\documentclass[twocolumn,prb,aps,10pt]{revtex4-1}
\usepackage{graphicx,epsf}

\begin{document}

\title{Magnetic effects on nonlinear mechanical properties of a suspended carbon nanotube}

\author{A. Nocera$^{1,2}$, C.A. Perroni$^{3,4}$, V. Marigliano Ramaglia$^{1,4}$, G. Cantele$^{3}$, and V. Cataudella$^{3,4}$ }

\affiliation{$^{1}$CNISM, UdR Napoli Universita', Monte
Sant�Angelo, Via Cintia, I-80126
Napoli, Italy\\
$^{2}$Dipartimento di Fisica E. Amaldi, Universita' di Roma Tre, Via della Vasca Navale 84, I-00146 Roma,Italy\\
$^{3}$CNR$-$SPIN, Monte Sant�Angelo, Via Cintia, I-80126
Napoli, Italy\\
$^{4}$Universita' degli Studi di Napoli Federico II, Complesso
Universitario Monte Sant'Angelo, Via Cintia, I-80126 Napoli,
Italy}

\begin{abstract}
We propose a microscopic model for a nanoelectromechanical system
made by a radio-frequency driven suspended carbon nanotube (CNT)
in the presence of an external magnetic field perpendicular to the
current. As a main result, we show that, when the device is driven
far from equilibrium, one can tune the CNT mechanical properties
by varying the external magnetic field. Indeed, the magnetic field
affects the CNT bending mode dynamics inducing an enhanced damping
as well as a noise term due to the electronic phase fluctuations.
The quality factor, as observed experimentally, exhibits a
quadratic dependence on external magnetic field strength. Finally,
CNT resonance frequencies as a function of gate voltage acquire,
increasing the magnetic field strength, a peculiar dip-peak
structure that should be experimentally observed.




\end{abstract}

\maketitle
\newpage

\section{Introduction}

NanoElectroMechanical Systems (NEMS)  made of suspended CNTs have
received increasing attention
recently\cite{Schmid,Steele,Huttel,Meerwaldt1,Bachtold,LeRoy,Sassetti1,Sassetti2}.
These devices are ideal for NEMS applications due to the extreme
mechanical properties of CNTs (low mass density and a high Young's
modulus), resulting in a wide range of resonance frequencies for
the fundamental bending mode vibration (from MHz up to
GHz\cite{Champagne} range). In particular, CNT-based
electromechanical devices working in the semiclassical regime
(resonator frequencies in MHz range compared to an electronic
hopping frequency from the leads of the order of tens of GHz) have
attracted great interest due to the extremely large quality
factors ($Q>10^{5}$) attainable\cite{Witkamp,Steele,Huttel}.
Furthemore, at cryogenic temperatures, such devices behave as
quantum dots\cite{Nygard,Bock,Dek} with a strong interplay between
single-electron tunneling and bending mode mechanical
motion\cite{Steele,Meerwaldt2,NoceraB,Labadze,Blanter,Blanter1}.
This means that the electronic current is very sensitive to the
CNT bending mode dynamics and can be used as a quantum (due to the
intrinsic quantum nature of the charge carriers) measurement
device. On the other hand, large quality factors allow to tune CNT
mechanical properties, e.g. the resonance frequency of the bending
mode, by adjusting electronic parameters such as gate and bias
voltages\cite{Steele,Meerwaldt2}. As a main consequence, one
expects that the application of a static magnetic field
perpendicular to the CNT device modifies the above picture since
the electronic current flow is affected by the Lorentz force.
This, in turn, induces a change in the mechanical properties of
the CNT resonator that can be measured through the current itself.

Motivated by recent experiments\cite{Schmid,Meerwaldt2}, we study
a general model describing the mechanical properties of a
radio-frequency driven suspended CNT-based NEMS in the presence of
a transverse magnetic field.
Indeed, we have recently shown that the effects of damping, spring
stiffening and softening, and nonlinearity observed in similar
devices in the absence of a magnetic
field\cite{Steele,Huttel,Meerwaldt} can be successfully described
in terms of a very simple effective model\cite{NoceraB}. In that
model, the CNT description is reduced to a single electronic
level\cite{Weick1,NoceraB} coupled to two metallic leads and
interacting by means of a charge-displacement term with a single
vibrational degree of freedom describing the bending mode. Due to
the low frequency associated to the bending mode\cite{Nota01}, the
vibrational CNT dynamics is described by a classical Langevin
equation\cite{Pisto,Nocera,Faz,Bode1,Bode2,Brandes,Brandes1,Bode1,Bode2,Weick1}.
In this simple approach, we were able to reproduce, in the absence
of magnetic field, the main results reported in
Ref.[\onlinecite{Steele}] and [\onlinecite{Huttel}] on a similar
CNT-resonator device and even to predict features not yet
observed. In particular, we have shown that when bias voltages are
smaller than the broadening due to tunnel coupling, the resonance
frequency shows a single dip as a function of gate voltage while,
at bias voltages exceeding the broadening due to tunnel coupling,
the resonance frequency shows a double dip structure.  The
successive experimental observation of the predicted
effects\cite{Meerwaldt2} has confirmed the validity of the model
adopted by us\cite{Nocera,NoceraB} in describing these devices.

The main result of the present paper is to prove that the external
magnetic field modifies the mechanical properties of the
CNT-resonator. In particular, the magnetic field provides an
additional damping mechanism for the resonator mechanical motion.
Interestingly, a quadratic decrease of the quality factor $Q$ as a
function of the external magnetic field strength, in quantitative
agreement with the experiment performed in
Ref.[\onlinecite{Schmid}], emerges.

When the device is driven far from the equilibrium we find that,
increasing the magnetic field,  the peculiar features (single or
double dip) observed in the resonance frequency against gate
voltage curves get distorted and acquire a dip-peak structure that
could be experimentally observed. In particular, this effect
should be more easily detectable at bias voltages that exceed the
broadening due to tunnel coupling.

We show that, at a fixed gate, bias voltage and temperature, if
charge and current variations of the opposite sign occur, damping
increases and $Q$ is reduced. Vice-versa, charge and current
variations of the same sign reduce damping with a consequent
increase of quality factors.


As concerns the physical mechanism triggered by the magnetic
field, we show that the application of a field perpendicular to
the current flux modifies all the terms describing the
CNT-resonator dynamics. Actually, the coupling with a transverse
magnetic field introduces an electronic tunneling phase which
depends on the mechanical displacement of the CNT-resonator
itself. This modifies the effective force acting on the resonator
by a pure non-equilibrium correction term proportional to the
magnetic field as well as to the electronic current\cite{Nota11}.
Moreover, even at zero bias voltage, damping and diffusive terms
are both modified by quantum electronic current-current
fluctuations corrections whose strength is quadratic in magnetic
field. Finally, we further show that, at zero bias,
charge-displacement and magnetic field mediated
electron-oscillator couplings cooperate behaving as a whole as a
standard thermal bath at leads' temperature.


The paper ends with a study of the device response when the
CNT-resonator motion is actuated by an external antenna at fixed
frequency and amplitude. In this case, the current-gate
voltage characteristic exhibits specific structures corresponding to the mechanical resonances (antenna frequency equal to the bending mode frequency).



The paper is organized as follows: In Sec. II we discuss the model,
In Sec. III we construct, by means of the adiabatic
approximation, the stochastic Langevin equation for the dynamics
of the oscillator including magnetic field  and
external antenna effects. In Sec. IV we present numerical results.

\section{Model}

\begin{figure}
\centering
{\includegraphics[width=8.0cm,height=3.0cm,angle=0]{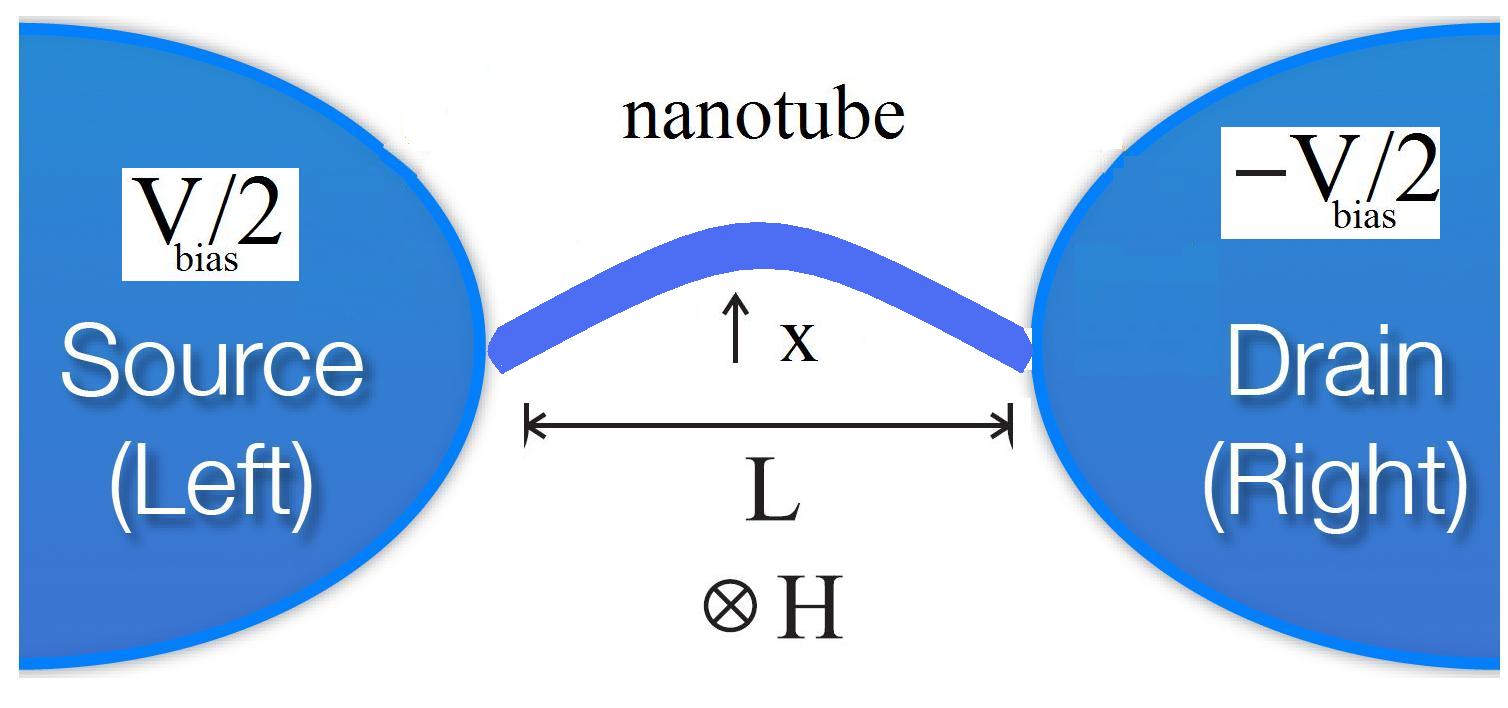}}
\caption{A carbon nanotube (CNT) subject to an external magnetic
field $H$ suspended between two normal metal leads biased by a
voltage $eV_{bias}$.}\label{fig0}
\end{figure}

We consider the system sketched in Fig.(\ref{fig0}), which shows a
single wall CNT suspended between two normal metal leads. An
external magnetic field $H$ is applied perpendicular to the CNT.
We also restrict the CNT mechanical degrees of freedom to the low
frequency bending mode and model it as a harmonic oscillator with
frequency $\omega_{0}$.

The electronic part of the CNT device is modeled as a single
electronic level coupled to the leads through standard tunneling
terms\cite{Nocera,NoceraB}. The electronic Hamiltonian is
\begin{equation}\label{Hel}
\hat{\cal H}_{el}=U_{gate}{\hat d^{\dag}}{\hat
d}+\sum_{k,\alpha}V_{k,\alpha}^{H}(x){\hat
c^{\dag}_{k,\alpha}}{\hat d}+ h.c.+
\sum_{k,\alpha}\varepsilon_{k,\alpha}{\hat
c^{\dag}_{k,\alpha}}{\hat c_{k,\alpha}},
\end{equation}
where CNT's electronic level has energy $U_{gate}$ with creation
(annihilation) operators ${\hat d^{\dag}} ({\hat d})$. The
operators ${\hat c^{\dag}_{k,\alpha}} ({\hat c}_{k,\alpha})$
create (annihilate) electrons with momentum $k$ and energy
$\varepsilon_{k,\alpha}=E_{k,\alpha}-\mu_{\alpha}$ in the left
($\alpha=L$) or right ($\alpha=R$) free metallic leads. The
chemical potentials in the leads $\mu_{L}$ and $\mu_{R}$ are
assumed to be biased by an external voltage
$eV_{bias}=\mu_{L}-\mu_{R}$.

In the presence of a magnetic field, the phases of the tunneling
amplitudes with the leads depend on the CNT displacement
$x$,\cite{Shekhter06,Radic}
\begin{eqnarray}\label{hoppingH}
V_{k,L}^{H}(x) &=& V_{k,L}e^{-\imath p x},\nonumber\\
V_{k,R}^{H}(x) &=& V_{k,R}e^{\imath p x},
\end{eqnarray}
where $p=\delta e H L/2 \hbar $ gives the CNT momentum change
induced by the Lorentz force when an electron tunnels from the CNT
to a lead, and $\delta\simeq1$ is a numerical factor determined by
the spacial profile of the fundamental mode\cite{PistoB}. Above,
$e$ is the modulus of the electron charge, $\hbar$ is the Plank
constant and $L$ is the CNT length. Slow time varying tunneling
amplitudes $V_{k}(t)$ are also relevant in the case of adiabatic
quantum pumping through quantum dots\cite{Faz,Citro}. Furthermore,
when the external magnetic field values are sufficiently small,
such as the Zeeman splitting is negligible compared to broadening
due to tunnel coupling, we can neglect the effect of the
electronic spin degrees of freedom (this issue will be considered
elsewhere\cite{NoceraSpin}).

For the sake of simplicity, we will suppose symmetric coupling
$V_{k,L}=V_{k,R}$ and a flat density of states for the leads
$\rho_{k,\alpha}\mapsto \rho_{\alpha}$, considered as thermostats
at finite temperature $T$, within the wide-band approximation
($V_{k,\alpha} \mapsto V_{\alpha}$, $\alpha=L,R$)
\cite{Weick,Weick1}.
Definitely, the total tunneling rate is
$\hbar\Gamma=\sum_{\alpha=L,R}\hbar\Gamma_{\alpha}$, with
$\Gamma_{\alpha}=2\pi\rho_{\alpha}|V_{k,\alpha}|^{2}/\hbar$.

The Hamiltonian of the mechanical degree of freedom is given by
\begin{equation}\label{Hosc}
\hat H_{osc}={\hat p^{2}\over 2m} + {1\over 2}m \omega^{2}_{0}\hat
x^{2},
\end{equation}
characterized by the frequency $\omega_{0}$ and the effective mass
$m$ ($k=m\omega_{0}^{2}$). The charge-displacement interaction is
provided by\cite{Weick,Armour}
\begin{equation}\label{Hint}
{\hat H}_{int}=\lambda \hat x{\hat n},
\end{equation}
where $\lambda$ is the charge-displacement coupling strength and
${\hat n}=\hat d^{\dag}d$ represents the electronic charge-density
on the CNT\cite{Nota3}. Definitely, the overall Hamiltonian is
\begin{equation}\label{Htot}
\hat{\cal H}=\hat{\cal H}_{el}+ \hat H_{osc} +{\hat H}_{int}.
\end{equation}

For the experiment discussed in Ref.[\onlinecite{Schmid}], one has
a strong separation between vibrational ($\omega_{0}\simeq 500$
MHz$\simeq$ 2$\mu$eV) and electronic time scales ($\Gamma \simeq
50$Ghz) so that we can solve our model in the adiabatic limit,
$\omega_{0}/\Gamma <<1$. The experimental values of bias voltages
and temperatures allow also a semi-classical treatment of the
oscillator dynamics\cite{Nocera,Mart,Pisto}. In this paper, we
will measure lengths in units of $x_{0}=r$, where $r$ is a small
fraction of CNT radius ($r=60pm$) appropriate to resolve the CNT
bending dynamics at relatively small temperatures ($T\simeq25mK$).
For the sake of simplicity, we will indicate dimensionless
displacement variable with $x$. Energies are measured in units of
$\hbar\Gamma=200\mu e V$, and times in units of
$t_{0}=1/\omega_{0}$. In terms of these units, the dimensionless
spring constant is $k/m\omega_{0}^{2} \simeq 1$, since, following
Ref.[\onlinecite{Schmid}], the effective mass of the nanotube is
$m=1.3 \times 10^{-21}kg$. Definitely, the adiabatic ratio is
$\omega_{0}/\Gamma=0.01$, while the dimensionless temperature
$k_{B}T=0.01$. Magnetic fields are measured in terms of the
quantity $B={H\over H_{0}}$ where our magnetic field unit is
$H_{0}=2\hbar/eLr \simeq 16.6 T$, since the CNT length is
$L\simeq700nm$. Throughout this paper, we keep fixed the
dimensionless charge-displacement coupling strength to
$\lambda=0.1$ (our force unit is $\hbar\Gamma/r$), corresponding
to an estimate $E_{p}=\lambda^{2}/2k\simeq 1\mu e V$, implying a
moderate coupling between the electronic and vibrational degrees
of freedom ($E_{p}/\hbar\omega_{0}=0.5$). Summarizing, the regime
of the relevant parameters is $\hbar\omega_{0}\simeq E_{p}\simeq
k_{B}T << eV^{eff}_{bias} \leq \hbar\Gamma$.


In the next section, we show how adiabatic approximation works on
the coupled electron-oscillator problem in the presence of a
transverse magnetic field.

\section{Adiabatic approximation}

As analyzed in the previous section, we work in the physical
regime where the vibrational motion of the CNT-resonator is 'slow'
with respect to all electronic energy scales and can be considered
"classical": $\omega_{0}<<\Gamma$. This regime of the parameters
leads to the adiabatic approximation for the electronic problem.
In contrast to previous works that treated the adiabatic
approximation in the absence of a magnetic
field\cite{Nocera,Faz,Bode1}, we here investigate the effect of a
transverse magnetic field on the electronic problem described by
the Hamiltonian Eq.(\ref{Htot}). We remark that the adiabatic
approximation has been used to describe larger
systems\cite{Nocera2} for the study of spectral and transport
properties of organic
semiconductors\cite{Nocera2,Nocera3,Nocera4}.

\subsection{Adiabatic approximation for the electron problem in the presence of a magnetic field}

In this subsection, we show how the adiabatic approximation on the
electronic CNT level Green function works in the presence of a
transverse magnetic field.

Assuming a slow time dependence of electronic Green functions
on the resonator displacement $x$, we are able to calculate
truncated expressions for the CNT level Green functions which
acquire a \`{}slow\'{} time dependence and, at first order, a
linear correction in the oscillator velocity. As a result of the
adiabatic approximation, the truncated CNT level Green functions
will depend on the instantaneous value of the position and
velocity of the resonator $G^{r,a,<,>}(\omega,x,\mbox{v})$.

The adiabatic expansion of the Fourier transformed retarded CNT
level Green function is
\begin{equation}\label{Grexp}
G^{r}(\omega,x,\mbox{v})=
G^{r}_{(0)}(\omega,x)+G^{r}_{(1)}(\omega,x,\mbox{v}),
\end{equation}
where the expression of $G^{r}_{(0)}(\omega,x)$ is
\begin{equation}
G^{r}_{(0)}(\omega,x)={1\over
\hbar\omega-U_{gate}(x)+\imath\hbar\Gamma/2},
\end{equation}
and that of $G^{r}_{(1)}(\omega,x,\mbox{v})$ is
\begin{equation}
G^{r}_{(1)}(\omega,x,\mbox{v})=\imath\hbar {\dot
U_{gate}(x)}G^{r}_{(0)}(\omega,x){\partial
G^{r}_{(0)}(\omega,x)\over
\partial\hbar\omega}.
\end{equation}
Above, $U_{gate}(x)=U_{gate}+\lambda x$ and the dot indicates the
time derivative ${\dot U_{gate}}=\lambda{\partial x\over
\partial t}=\lambda \mbox{v}$.

Using the adiabatic approximation\cite{Radic}
$x(t_{1})-x(t_{2})\simeq \dot x(t_{0}) (t_{1}-t_{2})$,
we obtain for the lesser and greater components in Fourier space
\begin{equation}\label{Siglesexp}
\Sigma^{<}_{leads}(\omega,\mbox{v})\simeq
\Sigma^{<}_{leads,(0)}(\omega)+\Sigma_{leads,(1)}(\omega,\mbox{v})
\end{equation}
where the expression of $\Sigma^{<}_{leads,(0)}(\omega)$ is
\begin{eqnarray}
\Sigma^{<}_{leads,(0)}(\omega)&=&\imath[\hbar\Gamma_{L}f_{L}(\omega)+\hbar\Gamma_{R}f_{R}(\omega)],
\end{eqnarray}
and that of $\Sigma_{leads,(1)}(\omega,\mbox{v})$ is
\begin{equation}\label{Sig1less}
\Sigma_{leads,(1)}(\omega,\mbox{v})=-\imath e {\tilde
H}\Bigg({\partial
[\hbar\Gamma_{L}f_{L}(\omega)+\hbar\Gamma_{R}f_{R}(\omega)] \over
\partial [eV_{bias}]}\Bigg)\mbox{v}.
\end{equation}
Above, we have defined ${\tilde H}=2p\hbar/e$. The adiabatic
correction to the lesser component of the leads self-energy in
Eq.(\ref{Sig1less}) is entirely due to the transverse magnetic
field and represent one of the main results of the present paper,
in contrast to previous works where magnetic field effects in the
adiabatic expansion were not
considered\cite{Mart,Pisto,Brandes,Nocera,NoceraB}.

Definitely, for the CNT level occupation we get
\begin{equation}
\langle \hat n\rangle (x,\mbox{v})\simeq \langle {\hat
n}\rangle_{(0)} (x)+\langle {\hat n}\rangle_{(1)}(x,\mbox{v}),
\end{equation}
where at zero order in the adiabatic expansion we get
\begin{equation}\label{occu0}
\langle {\hat n}\rangle_{(0)}(x)= \int{{d\hbar\omega\over 4\pi}
(f_{L}(\omega)+ f_{R}(\omega))C(\omega,x)},
\end{equation}
with the spectral function
$C(\omega,x)=-2\Im{G^{r}_{(0)}(\omega,x)}$ of the CNT level given
by
\begin{equation}\label{Spec0}
C(\omega,x)={\hbar\Gamma\over
(\hbar\omega-U_{gate}(x))^{2}+[\hbar\Gamma]^{2}/4}.
\end{equation}
The first order corrections in the adiabatic expansion are linear
in the oscillator velocity
\begin{equation}
\langle {\hat
n}\rangle_{(1)}(x,\mbox{v})=\mbox{v}[R_{(1)}(x)+R_{(2)}(x)],
\end{equation}
with
\begin{eqnarray}\label{R1x}
&&R_{(1)}(x) ={\lambda\over\Gamma}\int{{d\hbar\omega\over 2\pi}
g^{(1)}_{+}(\omega)C(\omega,x)T(\omega,x)},\\
&&R_{(2)}(x) = {e{\tilde H}\over2} \int{{d\hbar\omega\over 2\pi}
g^{(2)}_{+}(\omega)C(\omega,x)},
\end{eqnarray}
where we have defined the transmission function $T(\omega,x)$
\begin{equation}\label{Tfunction}
T(\omega,x)={\hbar\Gamma_{L}\hbar\Gamma_{R} \over
[(\hbar\omega-U_{gate}(x))^{2}+[\hbar\Gamma]^{2}/4]},
\end{equation}
and
\begin{eqnarray}\label{gomega}
g^{(1)}_{+}(\omega)&=&-{\partial
[f_{L}(\omega)+f_{R}(\omega)]\over
\partial \hbar\omega},\\
g^{(2)}_{+}(\omega)&=&-{\partial
[f_{L}(\omega)+f_{R}(\omega)]\over
\partial [eV_{bias}]}.
\end{eqnarray}
Above, $R_{(1)}(x)$ is the the adiabatic correction to the density
related to the charge-displacement coupling described in the
interaction Hamiltonian Eq.(\ref{Hint}), already described in many
papers in the literature\cite{Pisto,Brandes,Nocera}. $R_{(2)}(x)$
is the adiabatic correction exclusively due to magnetic-coupling
effects modifying the electronic phase of electrons flowing from
the leads to the CNT.

Finally, in the hypothesis of symmetric coupling to the leads
$\Gamma_{L}=\Gamma_{R}$, one can calculate the adiabatic expansion
for the symmetrized current $\langle {\hat I}\rangle=[\langle
{\hat I}_{L}\rangle-\langle {\hat I}_{R}\rangle]/2$
\begin{eqnarray}\label{tcurrent}
\langle {\hat I}\rangle (x,\mbox{v})&=& {e\over\hbar}
\int{d\hbar\omega\over 2\pi}
|G^{r}(\omega,x)|^{2}(\Sigma^{R,>}(\omega,\mbox{v})\Sigma^{L,<}(\omega,\mbox{v})\nonumber\\
&-&\Sigma^{L,>}(\omega,\mbox{v})\Sigma^{R,<}(\omega,\mbox{v})).
\end{eqnarray}
Using Eqs.(\ref{Grexp}),(\ref{Siglesexp}), we get
\begin{equation}
\langle {\hat I}\rangle (x,\mbox{v})\simeq \langle {\hat
I}\rangle_{(0)} (x)+\langle {\hat I}\rangle_{(1)}(x,\mbox{v}),
\end{equation}
where
\begin{eqnarray}\label{corr0}
\langle {\hat I}\rangle_{(0)}(x)=e\Gamma \int{d\hbar\omega\over
8\pi} (f_{L}(\omega)-f_{R}(\omega))C(\omega,x),
\end{eqnarray}
with linear corrections in the oscillator velocity
\begin{equation}
\langle {\hat
I}\rangle_{(1)}(x,\mbox{v})=\mbox{v}[U_{(1)}(x)+U_{(2)}(x)],
\end{equation}
with
\begin{eqnarray}
&&U_{(1)}(x) = -{e\lambda\over2} \int{d\hbar\omega\over
2\pi}g^{(1)}_{-}(\omega)C(\omega,x)T(\omega,x),
\end{eqnarray}
\begin{eqnarray}\label{U2}
&&U_{(2)}(x) = -{e^{2}\over\hbar}{\tilde H}\int{{d\hbar\omega\over
2\pi}g^{(2)}_{-}(\omega)T(\omega,x)},
\end{eqnarray}
where we have defined
\begin{eqnarray}\label{gomega}
g^{(1)}_{-}(\omega)&=&{\partial [f_{L}(\omega)-f_{R}(\omega)]\over
\partial \hbar\omega},\\
g^{(2)}_{-}(\omega)&=&{\partial [f_{L}(\omega)-f_{R}(\omega)]\over
\partial [eV_{bias}]}.
\end{eqnarray}
As already discussed referring to adiabatic corrections to the
average charge density of the CNT level, $U_{(1)}(x)$ is the the
adiabatic correction to the electronic current related to the
charge-displacement coupling described in the interaction
Hamiltonian Eq.(\ref{Hint}), already described in
Ref.[\onlinecite{Bode1}] and [\onlinecite{Bode2}]. $U_{(2)}(x)$ is
the adiabatic correction to the current exclusively due to
magnetic-coupling effects. As we show below, Eq.(\ref{U2}) is one
of the main result of the present paper, describing the increase
of damping acting on the CNT resonator given by the application of
a transverse magnetic field.

In next subsection, we show that, even in the presence of a
transverse magnetic field, the dynamics of the CNT-resonator can
be accurately described by a stochastic Langevin equation.

\subsection{Langevin equation for the oscillator}

In the absence of a magnetic field, the effect of the electron
bath and the charge-displacement coupling on the oscillator
dynamics gives rise to a stochastic Langevin equation with a
position dependent dissipation term and white noise force
\cite{Nocera}. As in Ref.[\onlinecite{NoceraB}], even in the
present case the equation for the oscillator dynamics can be
written as follows
\begin{eqnarray}\label{Langevin1}
m \ddot{x} &+& A(x)\dot{x}=F_{(0)}(x)+ \sqrt{D(x)}\xi(t)+A_{ext}\cos(\omega_{ext}t),\nonumber\\
\langle\xi(t)\rangle&=&0,\;\;\;\;\langle\xi(t)\xi(t')\rangle=\delta(t-t'),
\end{eqnarray}
where $\xi(t)$ is a standard white noise term. We have included in
our schematization the effect of an external antenna exciting the
motion of the CNT, where $A_{ext}$, $\omega_{ext}$ represent the
antenna amplitude and frequency, respectively. In this section, we
describe how all the terms appearing in above equation modify in
the presence of an external transverse magnetic field.

The total force acting on the CNT-resonator is 
\begin{equation}\label{for}
F=-k x-\lambda \langle {\hat n}\rangle(x,\mbox{v})+{\tilde H}
\langle {\hat I} \rangle(x,\mbox{v}).
\end{equation}
The linear elastic force exerted on the oscillator is modified by
two relevant \emph{nonlinear} correction terms: the former is
proportional to the electronic charge density on the CNT level
Eq.(\ref{occu0}), while the latter to the electronic current
Eq.(\ref{corr0}). The first term, due to the charge-displacement
interaction on CNT-resonator and proportional to $\lambda$ was
already discussed in Refs.[\onlinecite{Steele}] and
[\onlinecite{NoceraB}]. Far from equilibrium and in the presence
of a magnetic field, a magnetomotive coupling between the
CNT-resonator displacement and the electronic flow through the
device comes into play. Actually, the transverse magnetic field
introduces a phase in the electronic tunneling that is
proportional to the displacement of the CNT resonator as well as
on the field strength. This originates a Lorentz-like additive
correction, linear in the magnetic field strength and in the
electronic current, to the average force acting on the resonator.

In the limit of the adiabatic approximation, the force
Eq.(\ref{for}) can be decomposed in different expansion terms. It
explicitly depends on the oscillator position $x$ through
$U_{gate}(x)=U_{gate}+\lambda x$ and velocity $\mbox{v}$. The
force is
\begin{equation}\label{forMOD}
F(x,\mbox{v}) = F_{(0)}(x)+ F_{(1)}(x,\mbox{v}),
\end{equation}
where
\begin{equation}\label{for0}
F_{(0)}(x)=-kx -\lambda\langle {\hat n}\rangle_{(0)}(x)+{\tilde
H}\langle {\hat I}\rangle_{(0)}(x),
\end{equation}
and
\begin{eqnarray}
F_{(1)}(x,\mbox{v})&=&-\lambda\langle
{\hat n}\rangle_{(1)}(x,\mbox{v})+{\tilde H}\langle {\hat I}\rangle_{(1)}(x,\mbox{v})\nonumber\\
&=& -A(x)\mbox{v}.
\end{eqnarray}
The total damping term $A(x)$ is given by three contributions
\begin{equation}\label{damping}
A(x)=A^{\lambda}(x)+A^{H}(x)+A^{H,\lambda}(x),
\end{equation}
where both
\begin{equation}\label{dampinglambda}
A^{\lambda}(x)=\lambda R_{(1)}(x),
\end{equation}
coming from the charge-displacement coupling, and
\begin{equation}\label{dampingH}
A^{H}(x)=-{\tilde H} U_{(2)}(x),
\end{equation}
due to magnetic field coupling, are positive definite. The
function $A^{H,\lambda}(x)$ is proportional to both
charge-displacement coupling strength $\lambda$ and magnetic field
$H$
\begin{equation}\label{dampingHlambda}
A^{H,\lambda}(x)=\lambda R_{(2)}(x)-{\tilde H} U_{(1)}(x),
\end{equation}
and is not positive definite. Remarkably, we have verified that
the whole sum appearing in Eq.(\ref{damping}) is positive definite
in all parameters regime of our model. This shows that, using a
spinless fermionic model in the presence of normal (not
ferromagnetic) electronic leads, the CNT-resonator experiences no
negative damping regions. This is in contrast to results of
Ref.[\onlinecite{Radic}], where the authors use a normal and a
ferromagnetic lead and observe negative damping and consequent
nano-electromechanical self-excitations of the CNT-resonator
system.

A fluctuating term has to be be included to take correctly into
account the effect of the bath degrees of freedom.
When a magnetic field is present, the force-force fluctuations are
given by three contributions (see appendix A)
\begin{eqnarray}\label{fluforcetot}
\langle\delta\hat F(t)\delta\hat F(t')\rangle&=&
\lambda^{2}\langle\delta\hat n(t)\delta\hat n(t')\rangle
+ \nonumber\\
&-&{\tilde H}\lambda[\langle\delta\hat n(t)\delta\hat
I(t')\rangle+
\langle\delta\hat I(t)\delta\hat n(t')\rangle]+\nonumber\\
&+&{\tilde H}^{2}\langle\delta\hat I(t)\delta\hat I(t')\rangle,
\end{eqnarray}
where we get a mixed current-density fluctuation contribution
$[\langle\delta\hat n(t)\delta\hat I(t')\rangle+ \langle\delta\hat
I(t)\delta\hat n(t')\rangle]$, and a current-current fluctuation
contribution $\langle \delta I_{\alpha}(t)\delta I(t')\rangle$ to
the noise.

In the adiabatic limit, exploiting the effect of the 'fast'
electronic environment on the oscillator motion, one derives
\begin{equation}
\langle\delta\hat F(t)\delta\hat F(t')\rangle=D(x)\delta(t-t'),
\end{equation}
where in the presence of a magnetic field we have
\begin{equation}\label{diffusive}
D(x)=D^{\lambda}(x)+D^{H}(x)+D^{H,\lambda}(x),
\end{equation}
with
\begin{eqnarray}\label{diffusivel}
D^{\lambda}(x)&=&\lambda^{2}\hbar\int {d\hbar\omega\over
2\pi}G^{<}_{(0)}(\omega,x)G^{>}_{(0)}(\omega,x) =\nonumber\\
&=& \lambda^{2}\hbar \int{d\hbar\omega\over 2\pi} {
\hbar\Gamma_{L}f_{L}(\omega)+\hbar\Gamma_{R}f_{R}(\omega) \over
((\hbar\omega-U_{gate}(x))^{2}+[\hbar\Gamma]^{2}/4)^{2}}\times\nonumber\\
&\times&(\hbar\Gamma_{L}(1-f_{L}(\omega))+\hbar\Gamma_{R}(1-f_{R}(\omega)))
\end{eqnarray}
and
\begin{eqnarray}\label{diffusivelH}
&&D^{H\lambda}(x)= {e{\tilde H}\over2}\lambda\int
{d\hbar\omega\over
2\pi}|G^{r}_{(0)}(\omega,x)|^{2}C(\omega,x)\times\nonumber\\
&&\times(\Sigma^{L,>}_{(0)}(\omega)\Sigma^{L,<}_{(0)}(\omega)-\Sigma^{R,<}_{(0)}(\omega)\Sigma^{R,>}_{(0)}(\omega)) =\nonumber\\
&&= e\lambda{\tilde H} \int{d\omega\over 2\pi} {
\hbar\Gamma_{L}+\hbar\Gamma_{R} \over
[(\hbar\omega-U_{gate}(x))^{2}+[\hbar\Gamma]^{2}/4]^{2}}\times\nonumber\\
&&\times
\Big\{[\hbar\Gamma_{L}]^{2}f_{L}(\omega)(1-f_{L}(\omega))-[\hbar\Gamma_{R}]^{2}f_{R}(\omega)(1-f_{R}(\omega))\Big\}\nonumber\\
\end{eqnarray}
where $C(\omega,x)=-2\Im G^{r}_{(0)}(\omega,x)$ is the electronic
spectral function of the electronic level defined in
Eq.(\ref{Spec0}). The noise strength contribution coming from
current-current fluctuations is
\begin{eqnarray}\label{diffusiveH}
&&D^{H}(x)= {\tilde H}^{2}\hbar \int {d\hbar\omega\over
2\pi}\Big[f_{L}(\omega)-f_{R}(\omega)\Big]^{2}T(\omega,x)\times\nonumber\\
&&\times(1-T(\omega,x))+\Big\{f_{L}(\omega)(1-f_{L}(\omega))+\nonumber\\
&&+f_{R}(\omega)(1-f_{R}(\omega))\Big\}T(\omega,x),
\end{eqnarray}
where $T(\omega,x)={\hbar\Gamma\over4}C(\omega,x)$. In the absence
of electron bias voltage, one has $D(x)=2k_{B}TA(x)$, that is the
fluctuation-dissipation condition is verified for each fixed
position $x$. Moreover, it is possible to show that in the chosen
units, the dimensionless damping $A(x)$ (Eq.(\ref{damping})) and
diffusive term $D(x)$ (Eq.(\ref{diffusive})) result proportional
to the adiabatic ratio $\omega_{0}/\Gamma$.

\begin{figure}
\centering
{\includegraphics[width=9cm,height=8.0cm,angle=0]{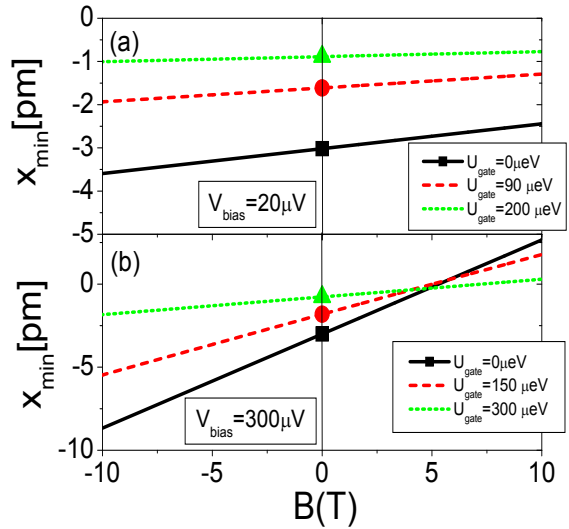}}
\caption{(Color online) Panel(a): minimum of the effective
potential (coming from the force Eq.(\ref{for0})) affecting the
CNT-resonator as a function of the magnetic field at low bias
$eV_{bias}=0.1\hbar\Gamma$ ($V_{bias}=20\mu V$ in our units).
Solid (black) line indicates $U_{gate}=0$, dashed (red) line
$U_{gate}=0.45$ ($V_{bias}=90\mu V$ in our units), dotted (green)
line $U_{gate}=1.0$ ($V_{bias}=200\mu V$ in our units). Panel(b):
same as above at large bias $eV_{bias}=1.5\hbar\Gamma$
($V_{bias}=300\mu V$ in our units). Solid (black) line indicates
$U_{gate}=0$, dashed (red) line $U_{gate}=0.75$ ($V_{bias}=150\mu
V$ in our units), dotted (green) line $U_{gate}=1.5$
($V_{bias}=300\mu V$ in our units).}\label{fig01}
\end{figure}

It is important to point out that, when there is no intrinsic
charge-displacement coupling ($\lambda=0$), in the absence of the
antenna effects and at zero bias ($V_{bias}=0$), the oscillator is
still governed by a Langevin equation
\begin{eqnarray}\label{Langevin1H}
m \ddot{x} &+& A^{H}(x)\dot{x}=kx+ \sqrt{D^{H}(x)}\xi(t),
\end{eqnarray}
with a harmonic force $F_{(0)}(x)=-kx$, an intrinsic
positive-definite dissipative term $A^{H}(x)$, and a diffusive
term $D^{H}(x)$ proportional to the thermal current-current noise.
Looking at Eqs.(\ref{dampingH}) and (\ref{diffusiveH}), one can
clearly see that a natural quadratic dependence of damping and
diffusive strength on the magnetic field emerges. This can be
explained observing that, even at zero bias voltage, the
electronic tunneling events, whose phase is dependent linearly on
the CNT displacements as well as on the magnetic field strength,
perturb the CNT mechanical motion with a force with zero average
(due to $\langle {\hat I}\rangle=0$, $\tilde H$ can be also
different from zero) and square mean proportional to the magnetic
field square. Definitely, even in the absence of external bias
voltage $V_{bias}$, the magnetic field applied perpendicular to
the CNT couples to the bending mode dynamics behaving as a
surrounding thermal bath at leads temperature $k_{B}T$.

\begin{figure}
\centering
{\includegraphics[width=9cm,height=8.0cm,angle=0]{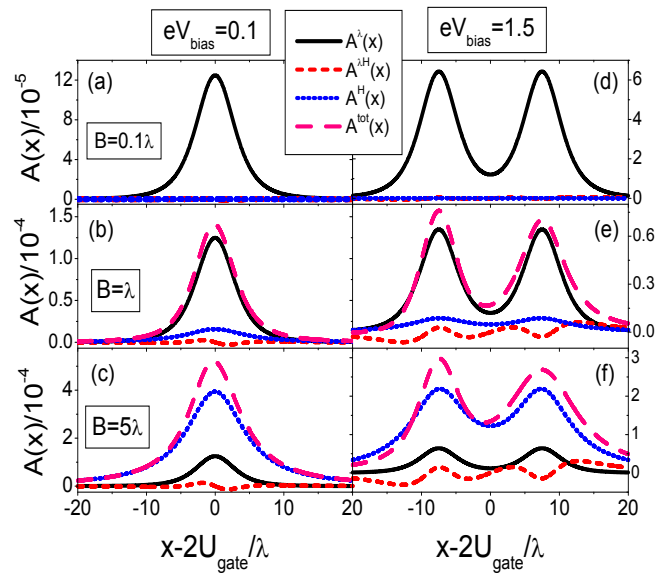}}
\caption{(Color online) Spatial dependence of the dimensionless
damping coefficient $A(x)$ at low bias (Panels(a-b-c)) and at
large bias voltage applied (Panels(d-e-f)). See main text for
discussion.}\label{fig1}
\end{figure}

\begin{figure}
\centering
{\includegraphics[width=9cm,height=8.0cm,angle=0]{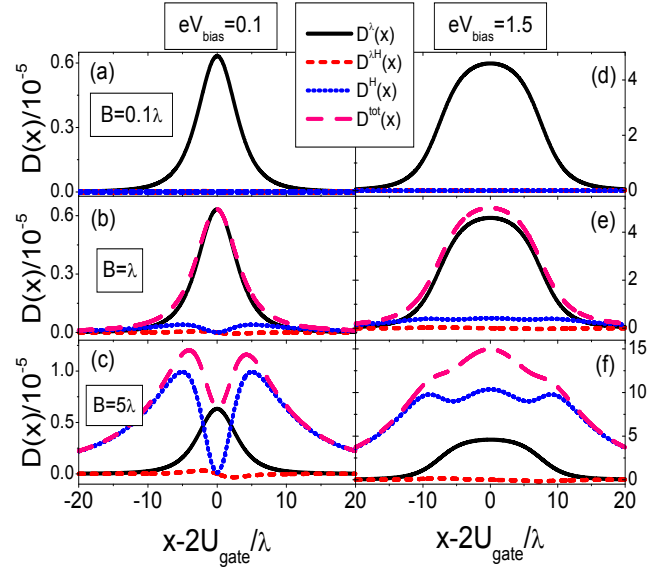}}
\caption{(Color online) Spatial dependence of the dimensionless
diffusive coefficient $D(x)$ at large bias voltage applied
(Panels(a-b-c)). See main text for discussion.}\label{fig2}
\end{figure}

We end this section with a systematic study of the spatial
dependence of the total force, the damping (see Fig.(\ref{fig1}))
and diffusive terms (see Fig.(\ref{fig2})) as a function of the
bias voltage as well as on the magnetic field.

As concerns the total force acting on the CNT resonator, we point
out that, for the magnetic field strengths investigated in this
paper, the effective potential preserves its parabolic shape with
a displaced minimum and renormalized curvature. For instance, when
a left-to-right current flows through the device (see the sketch
in Fig.(\ref{fig0})) in the presence of a positive magnetic field
(outgoing from the sketch reported in Fig.(\ref{fig0})), the
CNT-resonator effective potential minimum is displaced towards
positive displacements $x$ with respect to the minimum set by the
charge-displacement interaction (see Panel (a-b) of
Fig.(\ref{fig01})). In Fig.(\ref{fig01}), one can observe that,
the minimum of the effective potential acting on the resonator
depends linearly on the magnetic field strength. This comes from
the linear dependence on the magnetic field of the Lorentz-like
correction term to the force Eq.(\ref{for0}). In particular, as
shown in Panel(a) in Fig.(\ref{fig01}), in the low bias regime for
the device (small compared to the broadening of the CNT level),
the larger is the gate voltage, the smaller is the displacement of
the potential minimum as a function of the external magnetic field
with respect to the shift produced by the charge-displacement
interaction on the CNT (whose position is indicated by a (black)
square for $U_{gate}=0$, a (red) circle for $U_{gate}=0.45$
($V_{bias}=90\mu V$ in our units), and a (green) triangle
$U_{gate}=1.0$ ($V_{bias}=200\mu V$ in our units)). This can be
explained observing that in the low conducting regime of the
device the resonator is less effectively coupled with the
electronic subsystem. In the large bias regime (Panel(b) of
Fig.(\ref{fig01})), a smaller magnetic field is sufficient to
displace the potential minimum of the same quantity produced by
the sole charge-displacement interaction on the CNT. Again, the
larger is the gate voltage, the smaller is the displacement of the
potential minimum as a function of the external magnetic field
with respect to the shift produced by the charge-displacement
interaction on the CNT.

The renormalization of the effective potential curvature, that is
of the resonance frequency of the resonator, will be discussed in
subsection B of next section.

In this section, we limit ourself to discuss the damping term
$A(x)$, since for the diffusive term $D(x)$, unless explicitly
stated, a similar analysis can be done. As shown above (see
Eq.(\ref{damping})), we can distinguish between three
contributions to the friction affected by the oscillator: a
\emph{pure} charge-displacement contribution $A^{\lambda}(x)$,
depicted in Fig.(\ref{fig1}) with a solid (black) line, already
discussed in Refs.[\onlinecite{Nocera}] and
[\onlinecite{Brandes}]; a damping contribution due to
current-current fluctuations $A^{H}(x)$, depicted in
Fig.(\ref{fig1}) with a dotted (blue) line; a mixed damping term
due to current-density fluctuations (not positive definite),
indicated by $A^{\lambda H}(x)$ and depicted in Fig.(\ref{fig1})
with a short-dashed (red) line. The total damping $A(x)$ is
reported with a dashed (pink) line. As one can observe in
Panel(a), at low bias voltage, when the external magnetic field
strength is smaller than charge-displacement coupling $\lambda$,
the damping contributions coming from the current-current
$A^{H}(x)$ and current-density $A^{\lambda H}(x)$ fluctuations are
negligible with respect to that generated by the \emph{pure}
charge-displacement contribution $A^{\lambda}(x)$. In Panel (a),
$A^{\lambda}(x)$ and $A^{H}(x)$ have a single peak structure
centered at $x- 2U_{gate}/\lambda\simeq0$, while $A^{\lambda
H}(x)$ is an odd symmetric function with respect to this point. We
point out that these peculiar structures emerge only at larger
values of magnetic field (Panels(b-c) of Fig.(\ref{fig1})). The
total damping affecting the resonator is peaked at configurations
where large density variations take place $|x-2U_{gate}/\lambda| <
\hbar\Gamma/\lambda$. Indeed, the density of the CNT level goes
from a region $x-2U_{gate}/\lambda < -\hbar\Gamma/\lambda$
corresponding to almost completely filled states ($\langle
n\rangle\sim 1$) to a region $x-2U_{gate}/\lambda
> \hbar\Gamma/\lambda$ corresponding to completely empty states
($\langle {\hat n}\rangle\sim 0$). Definitely, the CNT level
experiences an unit charge variation across the
$|x-2U_{gate}/\lambda| < \hbar\Gamma/\lambda$
region\cite{NoceraB,Meerwaldt2}.

At large bias voltages applied (Panel (b)), $A^{\lambda}(x)$ has
two peaks centered at $x-2U_{gate}/\lambda\simeq
eV_{bias}/2\lambda$ and $x-2U_{gate}/\lambda\simeq
-eV_{bias}/2\lambda$, respectively. $A^{H}(x)$ shows the same
behavior, while $A^{\lambda H}(x)$ is an odd symmetric function
with respect to these two points. As in Panel (a), $A^{H}(x)$ and
$A^{\lambda H}(x)$ are negligible with respect to
$A^{\lambda}(x)$. The total damping affecting the resonator is
peaked at configurations where the CNT level experiences a
half-unit charge variation across the $|x-2U_{gate}/\lambda\pm
eV_{bias}/2\lambda| < \hbar\Gamma/\lambda$
regions\cite{NoceraB,Meerwaldt2}.

When the external magnetic field is turned on, an enhanced damping
as well as as noise strength emerges with a quadratic dependence
on the magnetic field intensity
Eqs.(\ref{damping})-(\ref{diffusive}). In Panels(b-e) of
Fig.(\ref{fig1}), one can observe that, as the dimensionless ratio
$B/\lambda$ is equal to one, the total damping affecting the
resonator is only slightly perturbed by the application of the
magnetic field. At low bias, $A(x)$ preserves its single peak
structure with an enhanced strength (dashed (pink) curve in Panel
(b) of Fig.(\ref{fig1})). At large bias, the strength of the two
peaks becomes asymmetric, with an enhanced damping of the peak at
$x-2U_{gate}/\lambda\simeq -eV_{bias}/2$. This effect can be
explained as follows: when a magnetic field is applied to the
device, the resonator starts to feel even the variations of the
electronic current flowing through the CNT as a function of the
gate voltage (see Eq.(\ref{for0})). These current variations are
positive for $x-2U_{gate}/\lambda<0$ and negative otherwise. At
$x-2U_{gate}/\lambda\simeq -eV_{bias}/2$, large negative
variations of the electronic density and positive variations of
the electronic current cooperate giving an enhanced damping.

We intend now to study the regime realized when the external
magnetic field strength is larger than charge-displacement
coupling strength $\lambda$. In this case, the contribution to the
damping coming from the current-current fluctuations $A^{H}(x)$
are dominant with respect to those corresponding to
charge-displacement $A^{\lambda}(x)$ and current-density
$A^{\lambda H}(x)$ fluctuations. In the low bias regime, the total
damping term preserves its single peak structure which, due to the
intrinsic asymmetry of the current-density term $A^{\lambda
H}(x)$, is slightly distorted. For the same reason, in the large
bias regime, the double dip structure of the total damping term is
preserved with an enhanced asymmetry. In the large magnetic field
regime, it is important to point out the particular spatial
dependence of the noise strength $D(x)$ (Panels (c-f) of
Fig.(\ref{fig2})). Here, the noise contribution due to the
current-current fluctuations ($D^{H}(x)$) emerges with the
characteristic double peak structure even at low bias regime
(dotted (blue) curve in Panel (c) of Fig.(\ref{fig2}) ). Comparing
the dashed (pink) curves in Panel (c) of
Figs.(\ref{fig1})-(\ref{fig2}), one can observe that, even at low
bias voltage, the application of a large magnetic field drive the
CNT-resonator far out of equilibrium, breaking the validity of the
Einstein relation $D(x)=2k_{B}T_{eff}A(x)$ with an effective
temperature. Far from equilibrium, this relation is strictly valid
only at very low bias voltages\cite{Nocera,Mart}.


In the next section, we study numerical results of our model
concerning mechanical properties of CNT-resonator (resonance
frequency and quality factor) as well as the electronic
observables inherent to the transport problem (I-V
characteristic).

\section{Mechanical and electronic characteristics of the device}

Given the assumption about the separation between the slow
vibrational and fast electronic (tunneling) timescales, the
problem of evaluating a generic observable (electronic or not) of
the system reduces to the evaluation of that quantity for a fixed
position $x$ and velocity $\mbox{v}$ of the oscillator, with the
consequent averaging over the stationary probability distribution
$P(x,\mbox{v})$. From the solution of the Langevin equation
(\ref{Langevin1}), one can determine the distribution
$P(x,\mbox{v})$ which allows to calculate all the electronic
observables $O$:
\begin{eqnarray}
\langle O\rangle &=& \int dxd\mbox{v} P(x,\mbox{v}) O(x,\mbox{v}).
\end{eqnarray}
We analyze in the next section the effects of the magnetic field
on the mechanical as well as electronic properties of the device.


\subsection{Device Quality factors}

One of the main findings of Ref.[\onlinecite{Schmid}] is the
observation of a quadratic dependence of the device quality factor
$Q$ on external magnetic field strength. Within our model, as also
stressed in the previous sections, such a quadratic dependence on
$B$ emerges naturally. In order to include back-actions effects of
the out of equilibrium electronic bath on the resonator, we have
calculated the average device quality factor as
\begin{equation}\label{Qfactor}
Q=\int_{-\infty}^{\infty}dx {1\over A(x)}P(x),
\end{equation}
where $A(x)$ is the total damping at a particular resonator
displacement $x$ and $P(x)$ is the reduced displacement
distribution probability of the CNT-resonator. We have verified
that this particular way of extracting quality factors is
completely equivalent to measure the width at half-high in the
current-frequency curves obtained in the linear response to an
external antenna exciting the nanotube motion\cite{NoceraB}.

\begin{figure}
\centering
{\includegraphics[width=9cm,height=8.0cm,angle=0]{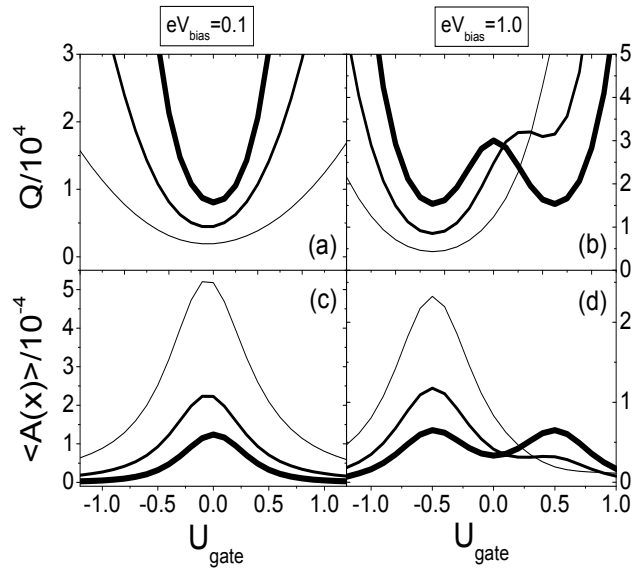}}
\caption{(Color online) Device quality factor as a function the
gate voltage $U_{gate}$ for different magnetic field strengths at
low (Panel (a)) and large (Panel (b)) bias voltage. Panels(c-d)
Same as above for the average total damping $\langle A(x)\rangle$
of the system. Panel(a-c): solid thick line $B=0.0$, solid
normal-thickness line $B=1.5$, and solid thin line $B=3.0$.
Panel(b-d): solid thick line $B=0.0$, solid normal-thickness line
$B=0.2$, and solid thin line $B=0.4$.}\label{fig6}
\end{figure}

Motivated by the experiment performed in Ref.[\onlinecite{Schmid}]
and by recent experimental study on a similar CNT
device\cite{Meerwaldt2}, we here performed a systematic study of
the quality calculated from our model as a function of the bias,
gate voltage as well as on the magnetic field. In Panels (a-b) of
Fig.(\ref{fig6}), we investigate the device quality factor $Q$ as
a function of gate voltage in the low and large bias voltage
regime, respectively. In the absence of a transverse magnetic
field, we reproduce the qualitative behavior obtained in the
experiment of Ref.[\onlinecite{Meerwaldt2}]. When bias voltages
are smaller than the broadening due to tunnel coupling, the
quality factor shows a single dip feature (solid (black) thick
line in Panel (a) of Fig.(\ref{fig6})). At bias voltages that
exceed (or are equal to) the broadening due to tunnel coupling,
the quality factor shows a double dip structure (solid (black)
thick line in Panel (b) of Fig.(\ref{fig6})).
This behavior, already addressed in Refs.[\onlinecite{Meerwaldt2}]
and [\onlinecite{NoceraB}], can be easily explained looking at the
average charge and dissipation of the CNT-resonator.
As also discussed referring to total damping affecting the
CNT-resonator in the previous section, at low bias voltage and in
the absence of magnetic field, the total average damping affecting
the resonator is peaked at electronic configurations where the CNT
level experiences an unit charge variation across the region where
the small conduction window is placed $|U_{gate}|<\hbar\Gamma$
(solid (black) thick line in Panel (c) of Fig.(\ref{fig6})). At
large bias voltages, the conduction window, whose extension is
proportional to $eV_{bias}$, becomes larger than the broadening of
the CNT level, so that the total average damping affecting the
resonator is peaked at electronic configurations where the CNT
level experiences half-unit charge variations, that is at
$|U_{gate}-eV_{bias}/2|<\hbar\Gamma$ and
$|U_{gate}+eV_{bias}/2|<\hbar\Gamma$. When the transverse magnetic
field is turned on, the above scenario modifies as follows. At low
bias voltages, the total damping affected by the CNT-resonator
increases quadratically with the field at every point in the
configuration space of the oscillator. Moreover, the CNT-resonator
distribution probabilities $P(x)$ depend slightly on the magnetic
field as well as on the gate voltages and are actually centered at
configurations close to the harmonic potential minimum $x\simeq0$
in the absence of charge-displacement interaction $\lambda$. The
overall result is an enhanced average total damping as one
increases the magnetic field (solid normal-thickness ($B=1.5$) and
thin ($B=3.0$) (black) lines in Panel (c) of Fig.(\ref{fig6})) and
a corresponding decrease of the quality factor in all the gate
voltage range investigated (solid normal-thickness ($B=1.5$) and
thin ($B=3.0$) (black) lines in Panel (a) of Fig.(\ref{fig6})). At
large bias voltage, the $P(x)$ still depends only slightly on the
magnetic field but is very spread on the configuration space.
Therefore, the average in Eq.(\ref{Qfactor}) reproduces the
spatial dependence structure of the total damping coefficient
reciprocal $1/A(x)$. The double peak structure of the average
total damping term (solid thick (black) line in Panel (d) of
Fig.(\ref{fig6})) is canceled by the magnetic field, giving a
single peak at $U_{gate}=eV_{bias}/2$ where a cooperation between
negative charge-density and positive current variations take place
(solid normal-thickness ($B=0.2$) and thin ($B=0.4$) (black) lines
in Panel (d) of Fig.(\ref{fig6})). As a consequence, the quality
factor loses its double dip structure getting a single dip at
$U_{gate}=-eV_{bias}/2$ (solid normal-thickness ($B=0.2$) and thin
($B=0.4$) (black) lines in Panel (b) of Fig.(\ref{fig6})).
\begin{figure}
\centering
{\includegraphics[width=9cm,height=8.0cm,angle=0]{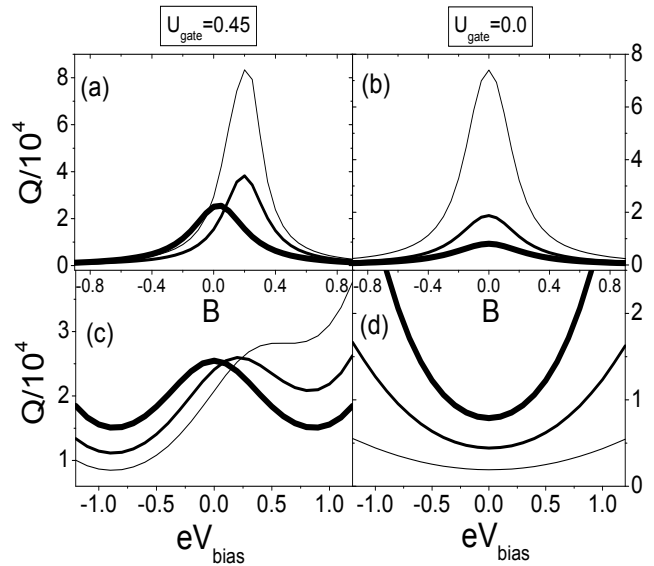}}
\caption{(Color online) Device quality factor as a function the
magnetic field strength $B$ for different bias voltages (Solid
thick line $eV_{bias}=0.1$, solid normal-thickness line
$eV_{bias}=0.75$, and solid thin line $eV_{bias}=1.5$) at low
(Panel (a)) and high (Panel (b)) conducting states. Panels(c-d)
Device quality factor as a function the bias voltage $eV_{bias}$
for different magnetic field strengths at low (Panel (c)) and high
(Panel (d)) conducting states. Panel(c): solid thick line $B=0.0$,
solid normal-thickness line $B=0.05$, and solid thin line $B=0.1$.
Panel(d): solid thick line $B=0.0$, solid normal-thickness line
$B=0.25$, and solid thin line $B=0.5$.}\label{fig7}
\end{figure}

\begin{figure}
\centering
{\includegraphics[width=9cm,height=8.0cm,angle=0]{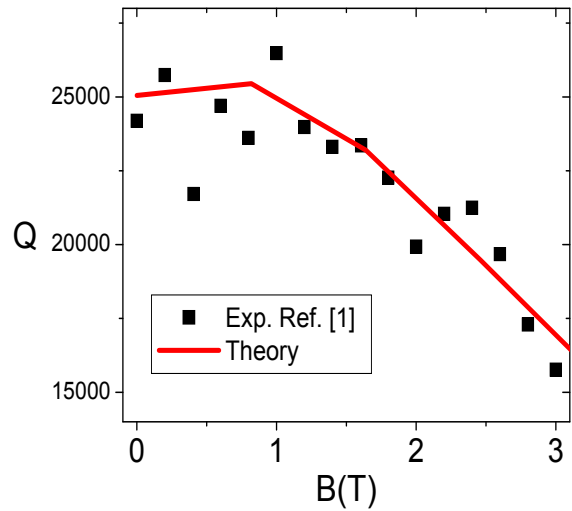}}
\caption{(Color online) Device quality factor as a function the
magnetic field strength. Squares represent experimental values
obtained in Ref.[\onlinecite{Schmid}] at $T=25mK$,
$V_{bias}=0.3mV$ and distance from the current peak
$U_{gate}=-90mV$. Solid (red) line is calculated $Q$ at
$k_{B}T=0.01$, $eV_{bias}=0.1$ and $U_{gate}=0.45$.}\label{fig5}
\end{figure}

We intend now to study the device quality factors as a function of
the transverse magnetic field $B$ comparing different conducting
states of the device. In Panel (a) of Fig.(\ref{fig7}), one can
observe calculated device quality factors as a function of the
magnetic field at a low conducting state of the device
($U_{gate}=0.45$). Different curves, from thicker to thinner,
refer to increasing bias voltages applied
$eV_{bias}=0.1-0.75-1.5$. At every fixed bias voltage, a clear
quadratic dependence of the total average damping on the magnetic
field strength is observed (not shown in Fig.(\ref{fig7})), with a
Lorentzian shape of the quality factor curves (see Panel (a) of
Fig.(\ref{fig7})). It is important to point out that the range of
magnetic field strengths experimentally investigated in
Ref.[\onlinecite{Schmid}], $B=0-3T$, corresponds to small magnetic
fields in our units (we recall that $H_{0}=16.6T$). Remarkably, at
low bias and small magnetic fields, a quadratic decrease of the
$Q$ against magnetic field is observed (see solid thick (black)
line in Panel (a) of Fig.(\ref{fig7})). In Fig.(\ref{fig5}), we
show the quantitative agreement between experimental and
calculated quality factors against magnetic field when the device
is in a low conducting state, with $eV_{bias}=0.1$ and
$U_{gate}=-0.45$. The slight increase of the quality factor $Q$ as
a function of the field for small magnetic fields, is due to
asymmetry introduced by the gate voltage $U_{gate}=0.45$ applied
to the device (see also Panel (a) of Fig.(\ref{fig7})). For gate
voltage equal to zero, that in the high conducting state of the
device, the calculated $Q$ against $B$ curve is a parabola with a
maximum at zero magnetic field applied.


Coming back to Panel (a) of Fig.(\ref{fig7}), one can observe an
interesting increase of the quality factor peak as a function of
the bias voltage. In particular, for $eV_{bias}=1.5$ (thinner line
in Panel (a) of Fig.(\ref{fig7})) a quality factor peak at
$B\simeq U_{gate}/2=0.225$ occurs. This can be directly related to
the average total damping dip, not shown in Fig.(\ref{fig7}). This
effect can be explained noting that, when the bias voltage applied
to the electronic device is increased, a transition from a single
peak to a double peak structure in the \emph{spatial} dependence
total damping affected by the CNT-resonator con be observed
(compare Panel(a) and (d) of Fig.(\ref{fig1})), while at the same
time, the displacement distribution probabilities $P(x)$ spread on
the configuration space remaining centered at configurations close
to the harmonic potential minimum $x\simeq0$ characteristic of the
low bias regime. The overall result is a reduction of the average
total damping affecting the CNT-resonator whose minimum is
translated by a quantity proportional to the gate voltage applied
to the device. This argument becomes even more clear when no gate
voltage is applied to the device which is therefore placed in a
high conducting state. In this case a perfect symmetry of Q-factor
curves with respect to zero magnetic field is obtained (see Panel
(b) of Fig.(\ref{fig7})).

We end this section with a study of the device quality factors as
function of the bias voltages and magnetic fields comparing low
and high conducting states of the device. In Panel (c) of
Fig.(\ref{fig7}), one can observe calculated device quality
factors as a function of the bias voltages at a low conducting
state of the device ($U_{gate}=0.45$). Different curves, from
thicker to thinner, refer to increasing magnetic field applied to
the device $B=0.0-0.05-0.1$. At zero magnetic field, a clear
double dip feature in the quality factor $Q$, as experimentally
observed in Ref.[\onlinecite{Meerwaldt2}], is visible. This can be
explained looking at the average total damping and in terms of the
average charge present on the CNT level. The total average
damping, in the absence of magnetic field, has two peaks at
$eV_{bias}=-2U_{gate}=-0.9$ and at $eV_{bias}=2U_{gate}=0.9$.
Indeed, as also discussed previously, the total average damping is
peaked at electronic configurations where the CNT level
experiences half-unit charge variations, that is at
$|U_{gate}-eV_{bias}/2|<\hbar\Gamma$ and
$|U_{gate}+eV_{bias}/2|<\hbar\Gamma$. Therefore, when the edges of
the conduction window (whose width is proportional to $eV_{bias}$)
meet the CNT level energy (given by $U_{gate}$), a maximum total
average damping (minimum quality factor) is observed. As also
discussed in reference of Fig.(\ref{fig6}), the double peak
structure of the average total damping term is canceled by the
magnetic field, giving a single peak at $eV_{bias}=-2U_{gate}$
where a cooperation between negative charge-density and positive
current variations take place. As a consequence, the quality
factor loses its double dip structure getting a single dip at
$eV_{bias}=-2U_{gate}$ (solid normal-thickness ($B=0.05$) and thin
($B=0.1$) (black) lines in Panel (c) of Fig.(\ref{fig7})).

In Panel (d) of Fig.(\ref{fig7}), we show calculated device
quality factors as a function of the bias voltages at a high
conducting state of the device ($U_{gate}=0.0$). Different curves,
from thicker to thinner, refer to increasing magnetic field
applied to the device $B=0.0-0.25-0.5$. As above, at zero magnetic
field, a single dip feature in the quality factor $Q$, as
experimentally observed in Ref.[\onlinecite{Meerwaldt2}], is
visible. This behavior can be discussed with the same argument
given for discussing the Panels(a-b) of Fig.(\ref{fig7}), where a
reduction of the total average damping as a function of the bias
voltage applied to the device was observed. Again, a decrease of
the quality factor in all the gate voltage range investigated as a
function of the magnetic field is observed (see Panel (d) of
Fig.(\ref{fig7})).

\subsection{Resonance frequency renormalization and current-voltage curves}

\begin{figure}
\centering
{\includegraphics[width=9cm,height=8.0cm,angle=0]{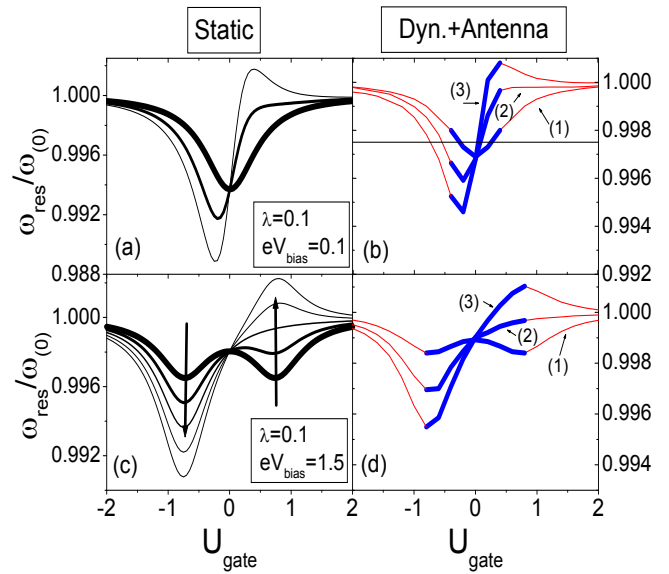}}
\caption{(Color online) Panel (a): Resonator frequency against
effective gate voltage calculated as minimum of the effective
potential in the static approximation at small bias
$eV_{bias}=0.1\hbar\Gamma$ for different magnetic field values:
Solid thick line $B=0.0$, solid normal-thickness line $B=1.5$,
solid thin line $B=3.0$. Panel (c): same as Panel (a) at large
bias $eV_{bias}=1.5\hbar\Gamma$ for different magnetic field
values: From thicker to thinner line $B=0.0-0.1-0.2-0.3-0.4$.
Panel (b-d): Resonator frequency calculated using an external
antenna (with $A_{ext}=10^{-3}$) at mechanical resonance against
effective gate voltage for same parameters of Panel (a-c),
respectively. Dashed (red online) and solid (blue online) portions
of each curve indicate resonance frequency values with positive
and negative current change $\Delta I$, respectively. In Panel
(d), only curves referring to magnetic field strengths
$B=0.0-0.2-0.4$ are reported.}\label{fig3}
\end{figure}

In this section, we address the magnetic field effects on the
renormalization of the CNT-resonator resonance frequencies and its
back-action effects on the current voltages curves of the device.
In order to study the CNT resonance frequency renormalization as
function of the gate voltage, we have compared results coming from
two ways of evaluation of the resonance frequencies. In the first
method, referred to as \emph{Static}, we evaluate the position of
the minima of the static potential arising from the generalized
force acting on the resonator (Eq.(\ref{for0})). In the second
method, referred to as \emph{Dynamic+antenna}, we have analyzed,
at every fixed value of the gate voltage, all the traces of
electronic current as a function of the antenna frequency
reporting with a red (blue) dot the resonance frequency values
with positive (negative) current change $\Delta I=I-I_{0}$ with
respect to background value $I_{0}$ obtained in the absence of the
antenna.

In Fig.(\ref{fig3}), we report the resonance frequencies of the
CNT-resonator as a function of the gate voltage comparing the two
methods outlined above. We address the low bias regime in Panels
(a-b), while the large bias regime is investigated in Panels(c-d).
In Panel (a) of Fig.(\ref{fig3}), different curves, from thicker
to thinner, refer to increasing magnetic field applied to the
device $B=0.0-1.5-3.0$. The same description was done in Panel
(c), where different curves refer to increasing magnetic field in
the range $B=0.0-0.2-0.4$. The thicker (black) lines in Panels (a)
and (c), corresponding to the absence of magnetic field, reproduce
qualitatively all results experimentally observed in
Ref.[\onlinecite{Meerwaldt2}]: when bias voltages are smaller than
the broadening due to tunnel coupling (Panel (a)), the resonance
frequency shows a single dip as a function of gate voltage. At
bias voltages that exceed the broadening due to tunnel coupling
(Panel (c)), the resonance frequency shows a double dip structure.
Actually, in this regime, the onset of a double dip structure was
already predicted by us in Ref.[\onlinecite{NoceraB}]. It is
important to point out that the resonance frequency
renormalization curves obtained in the presence of the external
antenna (Panel (b-d) of Fig.(\ref{fig3})) have the same
qualitative behavior (as a function of the gate) of those obtained
in the static approach. In the presence of an external antenna
with a finite amplitude, renormalization effects in the resonance
frequencies are less pronounced due to nonlinear
softening\cite{Steele,NoceraB}.

As already analyzed in Refs.[\onlinecite{Steele}] and
[\onlinecite{NoceraB}], when the device is in a low
current-carrying state, a peak in the current-frequency curve
signals the mechanical resonance (whose position is indicated by
thin (red) lines in Panels (b-d) of Fig.(\ref{fig3})), while in a
high current-carrying state, a dip in the current-frequency curves
is observed (whose position is indicated by thick (blue) lines in
Panels (b-d) of Fig.(\ref{fig3})). In the presence of a transverse
magnetic field, the different character of low and high conducting
states, signaled by a peak or a dip in current-frequency curves is
preserved (curves (2) and (3) in Panels (b-d) of
Fig.(\ref{fig3})).

\begin{figure}
\centering
{\includegraphics[width=9cm,height=8.0cm,angle=0]{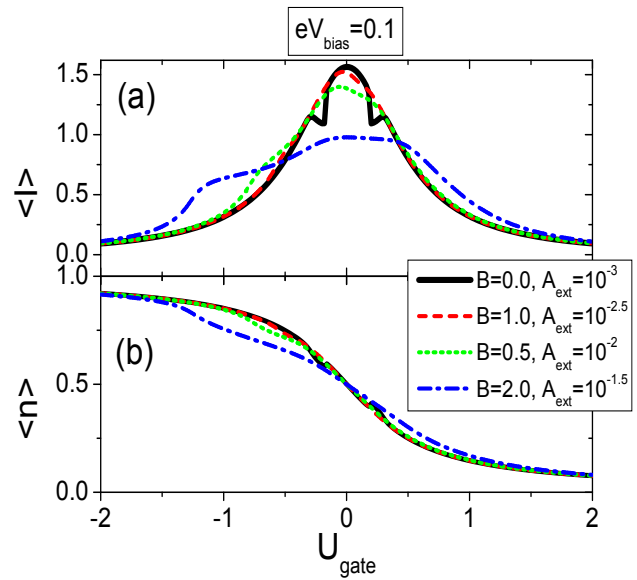}}
\caption{(Color online) Panel (a) Average electronic current
flowing through the CNT level at low bias ($eV_{bias}=0.1$) as
function of the gate voltage for different values of the magnetic
field and in the presence of a external antenna applied to the
device at fixed frequency $\omega_{ext}=0.9975$ and amplitude:
solid (black) line $A_{ext}=10^{-3}$, (dashed (red) line $B=1.0$
$A_{ext}=10^{-2.5}$, dotted (green) line $B=1.5$
$A_{ext}=10^{-2.0}$, dashed-dotted (blue) line $B=1.0$
$A_{ext}=10^{-2.5}$). Panel (b) Average electronic density on the
CNT level for the same parameter values as in Panel(a). See the
main text for detailed discussion.}\label{fig4}
\end{figure}

The peculiar features of CNT-resonator frequency renormalization
as a function of the gate can be explained with the same argument
used to describe the quality factors behavior in the previous
section. Indeed, in the absence of magnetic field, the resonator
frequency renormalization is maximum at electronic configurations
where the the CNT level experiences the largest charge-density
variations against the gate voltage,
\begin{equation}\label{keff}
k_{eff}=k+\lambda{\partial \langle {\hat n}\rangle\over\partial
U_{gate}}\Bigg|_{x=x_{min}}.
\end{equation}
Actually, at low bias voltages, an unit charge density variation
across the region where the small conduction window is placed
$|U_{gate}|<\hbar\Gamma$ (solid (black) thick line in Panel (a) of
Fig.(\ref{fig3})) occurs. At large bias voltages, the CNT
frequency renormalization is larger at electronic configurations
where the CNT level experiences a half-unit charge variation, that
is at $|U_{gate}-eV_{bias}/2|<\hbar\Gamma$ and at
$|U_{gate}+eV_{bias}/2|<\hbar\Gamma$.

When the transverse magnetic field is turned on, the above
scenario modifies as follows. The resonator frequency
renormalization is larger at electronic configurations where the
CNT level experiences the largest charge-density \emph{and}
current variations against the gate voltage,
\begin{equation}\label{keffH}
k_{eff}=k+\lambda{\partial \langle {\hat n}\rangle\over\partial
U_{gate}}\Bigg|_{x=x_{min}}-\lambda{\tilde H}{\partial \langle
{\hat I}\rangle\over\partial U_{gate}}\Bigg|_{x=x_{min}}.
\end{equation}
At low bias voltage, the single dip feature in the CNT-resonator
resonance frequency gets distorted (solid normal-thickness line in
Panel(a) of Fig.(\ref{fig3})) and acquires, in the limit of large
magnetic field (solid thin line in Panel(a) of Fig.(\ref{fig3})),
a dip-peak structure that could be experimentally observed.
Actually, the peak observed at $U_{gate}\simeq0.3$ corresponds to
an hardening of the CNT-resonator resonance frequency. This effect
can be explained as follows: when a magnetic field is applied to
the device, the resonator starts to feel even the variations of
the electronic current flowing through the CNT as a function of
the gate voltage (see Eq.(\ref{keffH})). These current variations
are positive for $U_{gate}<0$ and negative otherwise. At
$U_{gate}\simeq 0.3$, the positive (due to the positive sign of
the magnetic field) variations of the electronic current overcome
the negative variation of the electronic density giving a
hardening in the CNT resonance frequency. At $U_{gate}\simeq
-0.25$, one has negative variations of both density and current,
obtaining a more pronounced softening in the resonance frequency.
The effect outlined above is more pronounced in the large bias
regime (see Panel (c) of Fig.(\ref{fig3})). Here, the magnetic
field gives an enhanced softening dip at $U_{gate}\simeq
-eV_{bias}/2$ and an hardening peak at $U_{gate}\simeq
eV_{bias}/2$, where positive variations of the electronic current
cooperate with negative variation of the electronic density. In
both low and high bias regime, the hardening effect outlined above
could be experimentally observed.

The peculiar renormalization frequency effects discussed above
have a nontrivial back-action effects on the electronic density
and current-gate voltage characteristic of the device. In
Fig.(\ref{fig4}), we study the electronic CNT level density and
current as a function of the gate voltage in the presence of an
external antenna at fixed amplitude and frequency
$\omega_{ext}=0.9975$ (corresponding to the horizontal line in
Panel (b) of Fig.(\ref{fig3})) in the low bias regime of the
device\cite{Nota2}. When the external antenna frequency becomes
equal to the proper frequency of the resonator, we observe a dip
structure in both density and current at a gate voltage
corresponding to high conducting states of the device (solid
(black) line in Panels (a-b) of Fig.(\ref{fig4})). This feature,
that could be experimentally observed, is considered as a "dip"
with respect to corresponding curves in the absence of antenna or
with an antenna frequency far from the range of the proper
frequencies of the CNT-resonator (not shown in Fig.(\ref{fig4})).

When a transverse magnetic field is applied to the device, the CNT
frequency renormalization profile as a function of the gate
voltage changes (see Fig.(\ref{fig3})). Therefore, the mechanical
resonance condition between the external antenna frequency and the
proper frequency of the resonator occurs at different electronic
gate voltages. For sufficiently large magnetic fields, the
resonance can occur in correspondence of a low conducting state of
the device. As one can observe in Panels (a-b) of
Fig.(\ref{fig4}), a dip structure in the electronic density at a
more negative gate voltage and corresponding current peak (dotted
(green) and dashed-dotted (blue) lines in Panel (a-b) of
Fig.(\ref{fig4})) is visible. Actually, the above structures are
broadened due the reduction of the quality factor as a function of
the magnetic field. In the limit of very large magnetic fields, if
we keep fixed the amplitude of the external antenna, the fine
structures outlined above are completely washed out due to the
decrease of the device quality factors.

\section{Conclusions and Discussion}
In conclusion, we have studied a CNT-based electronic transistor
in the presence of an external magnetic field perpendicular to the
current flux. We were able to show that the application of a
transverse magnetic field modifies the bending mode CNT dynamics
giving an enhanced damping as well as a noise term originating
from the electronic phase fluctuations induced by the
displacements of CNT itself.

The effective force acting on the resonator is modified by a pure
nonequilibrium correction term proportional to the magnetic field
as well as to the electronic current flowing through the CNT.
Damping and diffusive terms are both modified by quantum
electronic current-current as well as density-current fluctuations
corrections whose strengths are quadratic and linear in the
magnetic field, respectively.

Within our model, a quadratic dependence of the device quality
factor $Q$ on external magnetic field strength, experimentally
observed in Ref.[\onlinecite{Schmid}], naturally emerges. This
behavior is understood in terms of a back-action of quantum
electronic current flow fluctuations on the bending mode dynamics.
A systematic study of device quality factor as a function of gate
and bias voltage in the presence of the magnetic field has also
been performed. All results are discussed observing the average
charge and electronic current variations with respect to gate
voltage applied to the device and can be summarized as follows. At
a fixed electronic conducting regime, if negative charge
variations and positive current variations occur, one has an
enhanced damping reducing the quality factor of the device.
Vice-versa, negative charge variations and negative current
variations reduce damping with a consequent increase of quality
factors.

We also show that, when the device is driven far from equilibrium,
one can tune CNT-resonator frequencies by varying the external
magnetic field: the peculiar (single or double dip) features in
the CNT-resonator resonance frequency, obtained in different
conducting regimes for the device, get distorted and acquire, in
the limit of large magnetic field, a peculiar dip-peak structure
that could be experimentally observed.

Finally, when the device is actuated by an external antenna at
fixed frequency and amplitude, the device current-gate voltage
response is modified by fine structure features any time the
mechanical resonance with the proper nanotube oscillation
frequency occurs. These structures can be tuned as a function of
the external field and could be experimentally observed. In this
sense, we have shown that, only exciting the CNT motion with
application of an external radio-frequency antenna, one can
observe a magnetic field dependence of the electronic current.

We point out that throughout this paper we do not take into
account of a magnetic field with a component longitudinal to the
CNT-resonator. This issue has been recently addressed in
Ref.[\onlinecite{Meerwaldt2}] and explained in terms of a more
sophisticated theoretical schematization of the CNT-resonator
electronic structure which has a cylindrical quasi-one dimensional
shape.

We end this section noting that it could be of outstanding
interest to study the possibility to include quantum corrections
to the oscillator dynamics as well as spin degrees of
freedom\cite{Radic} and electron-electron interaction effects in
the low bias regime. In particular, it has been recently proposed
to study the CNT bending mode dynamics by employing the spin-orbit
coupling between a single spin and nanomechanical
displacement\cite{Spettoesser,Flensberg,Perroni} in the presence
of a magnetic field. Quantum corrections becomes important when
the resonator and electronic time scales are of the same order on
magnitude. In this direction, it was shown in
Ref.[\onlinecite{Shekhter06}] that a magnetic field applied
perpendicular to the CNT results in negative magnetoconductance
due to quantum vibrations of the tube inducing an
Aharonov-Bohm-like effect\cite{Nota1} on the electrons crossing
the device. Work in this direction is in progress.


\section{ACKNOWLEDGMENTS}
A.Nocera acknowledges CNISM for the financial support. The
research leading to these results has received funding from the
FP7/2007-2013 under grant agreement N.264098 - MAMA.

\appendix
\section{Current-current and density-current fluctuations}
In this appendix, we illustrate how the calculation of the
force-force fluctuation (Eq.(\ref{fluforcetot})) can be performed
with the nonequilibrium Green function approach. In particular,
here we show how one can calculate in the adiabatic approximation
the current-current
\begin{equation}
S(t,t')=\langle \delta {\hat I}(t)\delta {\hat I}(t')\rangle
\end{equation}
and the density-current
\begin{equation}
M(t,t')=\langle [\delta {\hat n}(t)\delta {\hat I}(t')+ \delta
{\hat I}(t)\delta {\hat n}(t')]\rangle
\end{equation}
fluctuation terms appearing in Eq.(\ref{fluforcetot}). Recalling
that ${\hat I}=({\hat I}_{L}-{\hat I}_{R})/2$, it is easy to see
that $S(t,t')$ is made of three contributions
\begin{equation}
S(t,t')={1\over 2}[S_{L}(t,t')+S_{R}(t,t')+S_{LR}(t,t')],
\end{equation}
where
\begin{eqnarray}\label{Scurrent}
S_{L}(t,t')&=&\langle \delta {\hat I}_{L}(t)\delta {\hat
I}_{L}(t')\rangle,\\
S_{R}(t,t')&=&\langle \delta {\hat I}_{R}(t)\delta {\hat
I}_{R}(t')\rangle,\\
S_{LR}(t,t')&=&-\langle\{\delta {\hat I}_{L}(t),\delta {\hat
I}_{R}(t')\}\rangle.
\end{eqnarray}
The density-current fluctuation term $M(t,t')$ is given by
\begin{equation}
M(t,t')={1\over 2}[M_{R}(t,t')-M_{L}(t,t')],
\end{equation}
where
\begin{eqnarray}\label{Scurrent}
M_{L}(t,t')&=&\langle \{\delta {\hat I}_{L}(t),\delta {\hat
n}(t')\}\rangle,\\
M_{R}(t,t')&=&\langle \{\delta {\hat I}_{R}(t),\delta {\hat
n}(t')\}\rangle.\\
\end{eqnarray}
Above, $\{A,B\}=AB+BA$ is an anti-commutator. In this appendix we
limit to calculate $S_{L}(t,t')$ and $M_{L}(t,t')$ in the
adiabatic approximation, since for the other fluctuation terms the
derivation is similar.

We recall the expression for the current operator (through the
left barrier)\cite{Haug}
\begin{equation}\label{currentL}
I_{L}={\imath e\over \hbar}\sum_{k}
[V_{L,k}c_{k}^{\dag}d-V_{L,k}^{*}d^{\dag}c_{k}],
\end{equation}
We define $\delta {\hat I}_{L}(t)={\hat I}_{L}(t)-\langle {\hat
I}_{L}\rangle$, and plan to evaluate the correlation function (we
set $V_{L,k}=V_{k}$)
\begin{eqnarray}\label{noiset}
S_{L}(t,t') &=& {1\over 2} \langle\{\delta {\hat I}_{L}(t),\delta {\hat I}_{L}(t')\}\rangle \nonumber\\
&=& {1\over 2} \langle\{I_{L}(t),I_{L}(t')\}\rangle-\langle {\hat I}_{L}\rangle^{2}\nonumber\\
&=& {1\over 2}\Big({\imath e\over \hbar}\Big)^{2}\sum_{k,k'} \Bigg[V_{k}V_{k'}\langle c_{k}^{\dag}(t)d(t)c_{k'}^{\dag}(t')d(t')\rangle\nonumber\\
&-& V_{k}V^{*}_{k'}\langle c_{k}^{\dag}(t)d(t)d^{\dag}(t')c_{k'}(t')\rangle+\nonumber\\
&-&V^{*}_{k}V_{k'}\langle d^{\dag}(t)c_{k}(t)c_{k'}^{\dag}(t')d(t')\rangle+\nonumber\\
&+& V^{*}_{k}V^{*}_{k'} \langle
d^{\dag}(t)c_{k}(t)d^{\dag}(t')c_{k'}(t')\rangle\Bigg]\;+\; h.c. -
\langle {\hat I}_{L}\rangle^{2}.\nonumber\\
\end{eqnarray}
The Fourier transform of $S$ is called the noise spectrum; in what
follows we shall be particularly concerned with its zero-frequency
component, $S(0)=\int d(t-t') S(t-t')$ that is the relevant
quantity in the adiabatic expansion. In order to evaluate the
(nonequilibrium) expectation values occurring in Eq.(\ref{noiset})
in a systematic way, we first define the following contour-ordered
two-particle Green functions (we follow Ref.[\onlinecite{Haug}])
\begin{eqnarray}\label{2ndGreen}
&&G^{cd}_{1}(\tau,\tau')= \imath^{2} \langle T_{C} c_{k}^{\dag}(\tau)d(\tau)c_{k'}^{\dag}(\tau')d(\tau')\rangle \nonumber\\
&&G^{cd}_{2}(\tau,\tau')= \imath^{2} \langle T_{C} c_{k}^{\dag}(\tau)d(\tau)d^{\dag}(\tau')c_{k'}(\tau')\rangle \nonumber \\
&&G^{cd}_{3}(\tau,\tau')= \imath^{2} \langle T_{C} d^{\dag}(\tau)c_{k}(\tau)c_{k'}^{\dag}(\tau')d(\tau')\rangle \nonumber\\
&&G^{cd}_{4}(\tau,\tau')= \imath^{2} \langle T_{C}
d^{\dag}(\tau)c_{k}(\tau)d^{\dag}(\tau')c_{k'}(\tau')\rangle
\end{eqnarray}
The nonequilibrium noise correlator is then given by
\begin{eqnarray}\label{noiset1}
&&S_{L}(t,t')={1\over2}\Big({e\over\hbar}\Big)^{2}\sum_{k,k'} \Bigg[V_{k}V_{k'}G^{cd,>}_{1}(t,t') +\nonumber\\
&-& V_{k}V^{*}_{k'}G^{cd,>}_{2}(t,t')-
V^{*}_{k}V_{k'}G^{cd,>}_{3}(t,t') +\nonumber\\
&+&V^{*}_{k}V^{*}_{k'}G^{cd,>}_{4}(t,t') \Bigg]\;+\; h.c. -
\langle {\hat I}_{L}\rangle^{2},
\end{eqnarray}
where $G^{cd,>}_{i}(t,t')$ are the greater than components of the
contour-ordered counterparts $G^{cd}_{i}(\tau,\tau')$ defined in
Eq.(\ref{2ndGreen}).
In the adiabatic approximation, we consider the zero-order terms
of all Green's functions $G$. After lengthly but straightforward
calculations, starting from Eq.(\ref{noiset1}) we get (we follow
Ref.[\onlinecite{Haug}])
\begin{eqnarray}
S(t,t')&=&[S_{L}(t,t')+S_{R}(t,t')+S_{LR}(t,t')]/2\nonumber\\
&=&D^{H}(x)\delta(t-t'),
\end{eqnarray}
where $D^{H}(x)$ is given by Eq.(\ref{diffusiveH}) of the main
text. Eq.(\ref{diffusiveH}) is a well-known, and important result.
The first term accounts for thermal noise (i.e., it vanishes at
zero temperature), while the second term is a nonequilibrium term
(shot noise), which vanishes at zero bias.

\vskip1cm

For the mixed current-density contribution in the second line of
Eq.(\ref{fluforcetot}), we have (for the left lead)
\begin{eqnarray}\label{noiseMt}
&&M_{L}(t,t')= \langle\{\delta I_{L}(t),\delta n(t')\}\rangle =\nonumber\\
&&=\langle\{I_{L}(t),n(t')\}\rangle-2\langle {\hat I}_{L}\rangle^{2}\langle {\hat n}\rangle^{2}\nonumber\\
&&={\imath e\over \hbar}\sum_{k_{L}} \Bigg[V_{k_{L}}\langle c_{k_{L}}^{\dag}(t)d(t)d^{\dag}(t')d(t')\rangle\nonumber\\
&&-V^{*}_{k_{L}}\langle
d^{\dag}(t)c_{k_{L}}(t)d^{\dag}(t')d(t')\rangle\Bigg]\;+\; h.c. -
2\langle {\hat I}_{L}\rangle^{2}\langle {\hat n}\rangle^{2}.\nonumber\\
\end{eqnarray}
As for the the current noise spectrum $S$, in what follows we
shall be particularly concerned with the zero-frequency component
of $M_{L}(t,t')$, $M_{L}(\omega=0)=\int d(t-t') M_{L}(t-t')$ that
is the relevant quantity in the adiabatic expansion. In order to
evaluate the (nonequilibrium) expectation values occurring in
Eq.(\ref{noiseMt}) in a systematic way, we first define the
following contour-ordered two-particle Green functions
\begin{eqnarray}\label{2ndGreenM}
&&G^{Mcd}_{1,L}(\tau,\tau')= \imath^{2} \langle T_{C} c_{k_{L}}^{\dag}(\tau)d(\tau)d^{\dag}(\tau')d(\tau')\rangle, \nonumber\\
&&G^{Mcd}_{2,L}(\tau,\tau')= \imath^{2} \langle T_{C} d^{\dag}(\tau)c_{k_{L}}(\tau)d^{\dag}(\tau')d(\tau')\rangle. \nonumber \\
\end{eqnarray}
In terms of the previous Greens function in Eq.(\ref{2ndGreenM}),
The nonequilibrium current-density noise correlator $M$ is then
given by
\begin{eqnarray}\label{noiset1M}
&&M_{L}(t,t')={\imath e\over\hbar}\sum_{k_{L}} \Bigg[V_{k_{L}}G^{Mcd,>}_{1,L}(t,t') +\nonumber\\
&-& V^{*}_{k_{L}}G^{Mcd,>}_{2,L}(t,t')\Bigg]\;+\; h.c. -2\langle
I_{L}\rangle^{2}\langle {\hat n}\rangle^{2},\nonumber\\
\end{eqnarray}
where $G^{Mcd,>}_{i,L}(t,t')$ are the greater than components of
the contour-ordered counterparts $G^{Mcd}_{i,L}(\tau,\tau')$
defined in Eq.(\ref{2ndGreenM}). Following the same reasoning as
previous section one can show that
\begin{eqnarray}\label{IN}
&&M_{L}(t,t')\simeq {e\over \hbar}
\Bigg\{G^{>}(t,t')\Bigg[\int_{C} d\tau_{1}
G(t,\tau_{1})\Sigma_{L}(\tau_{1},t')\Bigg]^{<}\nonumber\\
&-&G^{<}(t',t)\Bigg[\int_{C} d\tau_{1}
\Sigma_{L}(t,\tau_{1})G(\tau_{1},t')\Bigg]^{>}+,\nonumber\\
&&+ G^{<}(t,t')\Bigg[\int_{C} d\tau_{1}
G(t,t_{1})\Sigma_{L}(\tau_{1},t')\Bigg]^{>}\nonumber\\
&-&G^{>}(t',t)\Bigg[\int_{C} d\tau_{1}
\Sigma_{L}(t,\tau_{1})G(\tau_{1},t')\Bigg]^{<}\Bigg\}.
\end{eqnarray}
where $\Sigma_{L}$ is the self-energy contribution due to the
coupling to the left lead and the integration is extended along
the Keldysh contour. The function,
\begin{equation}
f(t,t')=\Bigg[\int_{C} d\tau_{1}
G(t,\tau_{1})\Sigma_{L}(\tau_{1},t')\Bigg]^{<}
\end{equation}
can be calculated using Langreth's rules\cite{Haug}, giving
\begin{eqnarray}
f(t,t')&=&\int dt_{1}
G^{r}(t,t_{1})\Sigma^{<}_{L}(t_{1},t')\nonumber\\
&+&G^{<}(t,t_{1})\Sigma^{a}_{L}(t_{1},t'),
\end{eqnarray}
where $\Sigma^{a}_{L}$ is the advanced component of the left lead
self-energy. In the adiabatic approximation, we consider the
zero-order terms of all functions $G$ and $\Sigma$. After lengthly
but straightforward calculations, starting from Eq.(\ref{IN}) we
get
\begin{equation}
M(t,t')=[M_{R}(t,t')-M_{L}(t,t')]/2=D^{H\lambda}(x)\delta(t-t'),
\end{equation}
where $D^{H\lambda}(x)$ is given by Eq.(\ref{diffusivelH}).


\begin{thebibliography}{99}
\bibitem{Nota01} Here, the quantum regime effects predicted in
Refs.[\onlinecite{Shekhter06}] and [\onlinecite{Rastelli}] are not
relevant due to the very low ratio between the CNT bending mode
energy and the tunneling energy to the leads.
\bibitem{Nota11} In Ref.[\onlinecite{Jonson}], the authors
consider the presence of a Laplace-force analogous to that derived
by us and a cohesive force acting on the resonator in the presence
of an external transverse magnetic field. As suggested in
Ref.[\onlinecite{Jonson}], one can safely neglect the cohesive
force effects if $H>>l/L H_{0}^{*}$, where $H_{0}^{*}$ is a
characteristic magnetic field, $l$ is the distance of the quantum
dot from the leads, and $L$ is the CNT length. In our
schematization $l<<L$, since we reduce the entire CNT as a quantum
dot attached to the leads. Therefore, we can safely neglect the
cohesive force effects.
\bibitem{Nota3} Actually, more sophisticated approaches
involving the coupling with spatial fluctuations of the CNT
electronic density, recently proposed in Refs.
[\onlinecite{Sassetti1}] and [\onlinecite{Sassetti2}], are not
needed in the present case, while are very important to describe
Scanning Tunnel Microscopy experiments performed on suspended
CNTs\cite{LeRoy}.
\bibitem{Nota2} In the large bias regime, features of different origin of those investigated in this paper can
appear in the current-voltages characteristic of the device. These
switching effects are experimentally observed in dc-current
through the device investigated in Ref.[\onlinecite{Schmid}], and
have been identified as nano-electromechanical self-excitations of
the system, where positive feedback from single electron tunneling
drives mechanical motion. These features, smeared by the external
magnetic field, were predicted by a model introduced in Ref.
[\onlinecite{Blanter}] and [\onlinecite{Blanter1}] where energy
dependence of electronic tunneling amplitudes was also considered.
\bibitem{Nota1} When an electron crosses
the device, its wave function acquires an Aharonov-Bohm phase that
depends on the position of the displaced oscillator (see
fig.\ref{fig0} for a schematic representation of the system). The
total transmission results from the interference of all electronic
trajectories. Thus the Aharonov-Bohm phases generate a
magnetic-field dependence of the current that is absent, as we
show in this paper, in the case of a single classical path.
\end{thebibliography}
\end{document}